\documentclass{pasj00}

\SetRunningHead{K.~Saitou \etal}{NIR and X-Ray Observations of XSS\,J12270--4859}

\Received{2011/04/01}
\Accepted{2011/05/21}

\begin{document}
\title{Near-Infrared and X-Ray Observations of XSS\,J12270--4859}
\author{%
Kei~\textsc{Saitou},\altaffilmark{1,2}
Masahiro~\textsc{Tsujimoto},\altaffilmark{1}
Ken~\textsc{Ebisawa},\altaffilmark{1,2}
Manabu~\textsc{Ishida},\altaffilmark{1}
Koji~\textsc{Mukai},\altaffilmark{3}\\
Takahiro~\textsc{Nagayama},\altaffilmark{4}
Shogo~\textsc{Nishiyama},\altaffilmark{5}
and
Poshak~\textsc{Gandhi}\altaffilmark{1}
}
\altaffiltext{1}{%
Japan Aerospace Exploration Agency, Institute of Space and Astronautical Science,\\
3-1-1 Yoshinodai, Chuo-ku, Sagamihara, Kanagawa 252-5210}
\altaffiltext{2}{%
Department of Astronomy, Graduate School of Science, The University of Tokyo,
7-3-1 Hongo, Bunkyo-ku, Tokyo 113-0033}
\altaffiltext{3}{%
Code 662, NASA/Goddard Space Flight Center, Greenbelt, MD 20771, USA}
\altaffiltext{4}{%
Department of Physics, Graduate School of Science, Nagoya University,
Furo-cho, Nagoya 464-8602}
\altaffiltext{5}{%
Extra-Solar Planet Detection Project Office, National Astronomical Observatory of Japan,\\
2-21-1 Osawa, Mitaka, Tokyo 181-8588}
\email{ksaitou@astro.isas.jaxa.jp}
\KeyWords{%
infrared: stars ---
stars: individual (XSS\,J12270--4859) ---
stars: variables: other ---
X-rays: stars
}
\maketitle

\begin{abstract}
  XSS\,J12270--4859 (J12270) is an enigmatic source of unknown nature. Previous studies
  revealed that the source has unusual X-ray temporal characteristics, including
  repetitive short-term flares followed by spectral hardening, non-periodic dips, and
  dichotomy in activity; i.e. intervals filled with flares and those without. Together
  with a power-law X-ray spectrum, it is suggested to be a low-mass X-ray binary
  (LMXB). In order to better understand the object, we present the results of our
  near-infrared (NIR) photometry and linear polarimetry observations as well as X-ray
  spectroscopy observations, which overlap with each other partially in time, taken
  respectively with the InfraRed Survey Facility (IRSF) and the Rossi X-ray Timing
  Explorer (RXTE). We detected several simultaneous NIR and X-ray flares for the first
  time. No significant NIR polarization was obtained. We assembled data taken with IRSF,
  RXTE, Suzaku, Swift, and other missions in the literature and compared the flare
  profile and the spectral energy distribution (SED) with some representative high-energy
  sources. Based on some similarities of the repetitive NIR and X-ray flaring
  characteristics and the broad SED, we argue that J12270 is reminiscent of microquasars
  with a synchrotron jet, which is at a very low luminosity state of $\approx$10$^{-4}$
  Eddington luminosity for a stellar mass black hole or neutron star at a reference
  distance of 1~kpc.
\end{abstract}

\section{Introduction}\label{s1}
XSS\,J12270--4859 (hereafter J12270) is one of the most mysterious X-ray objects among
recent discoveries, which shows quite anomalous temporal characteristics both in the
X-ray and optical regimes \citep{saitou09,demartino10,pretorius09}. A possible
association with the Fermi $\gamma$-ray source 1FGL\,J1227.9--4852 has also been
proposed \citep{demartino10,stephen10,hill11}. It is conceivably a low mass X-ray binary
(LMXB) from its rapid flux variation and a power-law spectrum in the X-ray
\citep{saitou09,demartino10}, but conventional views of LMXBs are insufficient to
explain its unusual features such as a broad spectral energy distribution (SED) up to
GeV and anomalous X-ray variability.

To better understand the system, near-infrared (NIR) observations are key. Several
spectral components are expected in the NIR regime for LMXBs, which include the thermal
emission from the secondary star and outer part of the accretion disk, and non-thermal
cyclo-synchrotron emission by electrons accelerated in a jet or hot inner accretion
flow.  These components have a different dependence on wavelengths and time variability
(or lack thereof), thus comparison with NIR and other wavelength observations
\citep{pretorius09,saitou09,demartino10} will help decompose the broad
SED. Nevertheless, no results of targeted NIR observations have been presented to date
for this source besides all-sky surveys.

\medskip

The main purpose of this paper is to present the result of NIR photometry and linear
polarimetry observations using the InfraRed Survey Facility (IRSF) 1.4~m
telescope. During a part of our NIR observations, we also conducted simultaneous X-ray
observations using the Rossi X-ray Timing Explorer (RXTE). In addition, we retrieve the
unpublished archived X-ray data by the Swift satellite to complement the previous
reports by \citet{saitou09} and \citet{demartino10} showing Suzaku, XMM-Newton, and
RXTE data, and examine if these anomalous X-ray characteristics are common in this object.

The plan of this paper is as follows. In \S2, we briefly review the previous
observational findings of this object. In \S3 and \S4, we describe the observations and
analysis of the data separately for each observing facility. In \S5, we combine all the
results and discuss the possible nature of the system. The main results are summarized 
in \S6.

\section{Object Properties}\label{s2}
J12270 was initially discovered in the RXTE slew survey \citep{revnivtsev04} and later
identified as a hard X-ray emitter by INTEGRAL (\cite{bird07}; \cite{bird10}).  J12270
was once classified as an intermediate polar (IP), a binary system of a moderately
magnetized ($\sim 10^{5-7}$~G) white dwarf and a late-type companion based on optical
spectroscopy \citep{masetti06} and a hint of 860~s X-ray periodicity \citep{butters08}.

The results of X-ray spectroscopy \citep{saitou09,demartino10}, however, argue against
the IP nature of this source. The X-ray spectrum is featureless, not showing signature
of Fe emission lines, contrary to the defining IP characteristic of strong Fe emission
lines \citep{ezuka99}.

Anomalous X-ray temporal characteristics of J12270 were first discovered by
\citet{saitou09} using the Suzaku X-ray satellite and later confirmed by
\citet{demartino10} using the XMM-Newton satellite. The features include: (1)~repetitive
flux increase lasting for a short duration of a few hundred seconds (flares), (2)~sudden
flux decrease with no apparent periodicity (dips), (3)~spectral hardening after some
flux increase (hardening), and (4)~dichotomy in X-ray activity; i.e. intervals filled
with flares and those without. \citet{demartino10} also pointed out that there are two
types of dips; i.e., dips with hardening only seen immediately after flares and those
without hardening seen during quiescent phases without apparent periodicity.

Unique temporal characteristics were also reported in the optical and in the ultraviolet (UV)
regimes. \citet{pretorius09} found flux variation of a large amplitude ($>1$~mag), which
is accompanied by flickering and a distinctive pattern of variation. In optical
spectroscopy, they also reported that the equivalent width of the H$\alpha$ emission
line was twice as large as that reported by \citet{masetti06}. Using the broad-band
capability of XMM-Newton by combining the optical monitor (OM) and the X-ray CCD detectors,
\citet{demartino10} accumulated many new results revealing intriguing features of this
source: (1)~optical/UV flares are seen, which are always accompanied by X-ray flares,
(2)~X-ray dips in the quiescent phase are not seen in the UV, and (3)~a possible orbital
period of 4.32~hour in $V$ band photometry.

\citet{hill11} presented a detailed temporal and spectral analysis of the Fermi source,
for which they claim J12270 to be the most likely counterpart. The 0.1--300~GeV light
curve is stable with a 4-day time binning over 1 year. The 0.1--300~GeV luminosity is
$\sim$4.9$\times$10$^{33}$~erg~s$^{-1}$ at an assumed distance of 1~kpc, which is
comparable to that of X-ray luminosity in 0.2--12~keV band \citep{saitou09}. The
spectrum is represented by a power-law model with a photon index of 2.5 and shows a
significant cut-off at 4.1~GeV. Based on these properties, \citet{hill11} claimed that
the Fermi source is a binary millisecond pulsar.

\citet{hill11} also conducted radio imaging observations around the Fermi source using
the Australian Telescope Compact Array (ATCA) and the Giant Metrewave Radio Telescope
and found a radio counterpart to J12270. A significant detection was obtained at 5.5 and
9~GHz with a flux of 0.18 and 0.14~mJy, respectively. No millisecond pulsation was detected 
by follow-up observations using the Parkes telescope.

\section{Observations and Data Reduction}\label{s3}

\subsection{IRSF}\label{s3-1}
We had two runs of NIR observations in 2009 January and March with a total of 13~nights
(table~\ref{t1}). We used IRSF 1.4~m telescope in the
South African Astronomical Observatory. A part of the observations on March 13 and 15
was coordinated to be simultaneous with RXTE (\S~\ref{s3-2}). The NIR camera
at the Cassegrain focus of the telescope has two modes of operation: photometry with the
Simultaneous three-color InfraRed Imager for Unbiased Survey (SIRIUS:
\cite{nagashima99,nagayama03}) and linear polarimetry with the SIRIUS Polarimetry mode
(SIRPOL: \cite{kandori06}).

SIRIUS yields simultaneous $J$ (1.25~\micron), $H$ (1.63~\micron), and $K_{\textrm{s}}$
(2.14~\micron) band images using two dichroic mirrors.  A very fast readout (5~s per
frame) enables us to follow rapid flux changes of J12270 reported in the optical and
X-rays. SIRIUS has $1024\times 1024$~pixels with a pixel scale of 0\farcs45
pixel$^{-1}$, making a $7\farcm7\times 7\farcm7$ field of view (FoV).

SIRPOL is equipped with an achromatic (1--2.5~\micron) wave plate rotator unit over
SIRIUS and takes simultaneous $J$, $H$, and $K_{\mathrm{s}}$ band images at four
position angles (\timeform{0D.0}, \timeform{45D.0}, \timeform{22D.5}, and
\timeform{67D.5}) for measurements of linear polarization.  The Stokes parameters $I$,
$Q$, and $U$ are obtained as $I=\left(
I_{\timeform{0D.0}}+I_{\timeform{45D.0}}+I_{\timeform{22D.5}} +I_{\timeform{67D.5}}
\right) /2$, $Q=I_{\timeform{0D.0}}-I_{\timeform{45D.5}}$, and
$U=I_{\timeform{22D.5}}-I_{\timeform{67D.5}}$, where $I_{x}$ is the intensity of an
object at the position angle $x$. The polarization degree $P$ and the polarization angle
$\theta$ are then calculated as $P=\sqrt{Q^{2}+U^{2}}/I$ and $\theta =1/2 \arctan \left(
U/Q\right)$.

All SIRIUS and SIRPOL observations were carried out with a 15~s exposure per frame with
a 10-point dithering. The observing conditions were photometric with a typical seeing of
1\farcs4. Dark frames were obtained at the end of each night, while twilight flat frames
were obtained daily before and after the observation and averaged over a month.

We reduced the data following the standard pipeline procedure using the Imaging
Reduction and Analysis Facility (IRAF) software package. The procedure includes
subtracting dark frames, flat-fielding, masking bad pixels, and eliminating fringe
patterns by OH emission and the reset anomaly of the detector. Ten dithered frames were
then merged into one image. A small number of frames were discarded during the automated
pipeline processing due to incompleteness of the data.

\subsection{RXTE}\label{s3-2}
We observed J12270 using RXTE \citep{bradt93} twice on 2009 March 13 and 15 as a
target of opportunity program to conduct simultaneous X-ray and NIR observations
(table~\ref{t1}). Among three instruments onboard the satellite, we used the
Proportional Counter Array (PCA: \cite{jahoda06}), which consists of five Proportional
Counter Units (PCUs). The PCA, which is a non-imaging detector with a $1\arcdeg \times
1\arcdeg$ FoV, covers an energy range of 2--60~keV and has a total area of
6250~cm$^{2}$.

We used the data obtained by the top layer of PCU~2 with a time resolution of
16~s. Following the standard prescription, we discarded events when the separation
between the object and the satellite pointing was larger than $\timeform{0D.02}$ and the
Earth elevation angle was below 10$\arcdeg$. We further removed events during South
Atlantic Anomaly passages and 30~min thereafter, those within 600~s before and 150~s
after PCA breakdowns, and those when the electron to X-ray event ratio was larger than
0.1. As a result, the total exposure time was 16~ks (table \ref{t1}). The source events
are subject to contamination by other objects due to the lack of imaging capability and
a large FoV of the instrument. We estimate the contamination to be $\sim$10\% by
comparing the J12270 flux between RXTE and Suzaku \citep{saitou09}. We used the
HEASoft\footnote{See http://heasarc.gsfc.nasa.gov/docs/software/lheasoft/ for details.}
version 6.6 for the reduction of the RXTE data.

\subsection{Swift}\label{s3-3}
The Swift satellite \citep{gehrels04} observed J12270 twice on 2005 September 15 and 24
(table~\ref{t1}). The X-Ray Telescope (XRT: \cite{burrows05}) is equipped with an X-ray
CCD device for imaging-spectroscopy, which has a sensitivity in an energy range of
0.2--10~keV, an energy resolution of 140~eV at 5.9~keV, an effective area of
125~cm$^{2}$ at 1.5~keV, and a FoV of $23\farcm6 \times 23\farcm6$. We used the photon
counting mode with a frame time of 2.5~s.

The data were retrieved from the archive and reduced with the standard pipeline process,
yielding a total exposure time of $\sim$7~ks (table \ref{t1}). We used the HEASoft 
version 6.5 for the reduction of the Swift data.

\begin{table}
 \begin{center}
  \caption{Log of the observations}
  \label{t1}
  \begin{tabular}{lcccc}
   \hline
   Instrument  & Sequence & \multicolumn{2}{c}{Obs. start (UT)} & $t_{\rm{exp}}$\footnotemark[$\ddagger$] \\
               & number   & Date\footnotemark[$*$] & Time\footnotemark[$\dagger$] & (ks) \\
   \hline
   \multicolumn{5}{c}{\textit{IRSF}}\\
   \hline
   SIRIUS & ---      & 09-01-15 & 23:36 & \phantom{0}0.9 \\
          & ---      & 09-01-16 & 22:33 & \phantom{0}0.5 \\
          & ---      & 09-01-18 & 00:08 & \phantom{0}0.6 \\
          & ---      & 09-01-19 & 00:32 & \phantom{0}0.6 \\
          & ---      & 09-01-20 & 00:02 & \phantom{0}0.6 \\
          & ---      & 09-01-20 & 23:15 & \phantom{0}0.6 \\
          & ---      & 09-01-23 & 00:52 & \phantom{0}0.5 \\
          & ---      & 09-03-12 & 01:25 & \phantom{0}4.1 \\
          & ---      & 09-03-13 & 18:14 & 14.4 \\
          & ---      & 09-03-17 & 17:58 & 20.0 \\
   \hline
   SIRPOL & ---      & 09-03-14 & 23:35 & \phantom{0}4.2 \\
          & ---      & 09-03-15 & 18:38 & \phantom{0}8.4 \\
          & ---      & 09-03-16 & 18:38 & \phantom{0}8.4 \\
   \hline
   \multicolumn{5}{c}{\textit{RXTE}}\\
   \hline
   PCA    & 94416-01-01 & 09-03-13 & 20:06 & \phantom{0}9.7 \\
          &             & 09-03-15 & 20:42 & \phantom{0}6.5 \\
   \hline
   \multicolumn{5}{c}{\textit{Swift}}\\
   \hline
   XRT    & 0035101001 & 05-09-15 & 00:22 & \phantom{0}4.9 \\
          & 0035101002 & 05-09-24 & 06:09 & \phantom{0}1.9 \\
   \hline
   \multicolumn{5}{@{}l@{}}{\hbox to 0pt{\parbox{170mm}{
   \footnotesize
   \par\noindent
   \footnotemark[$*$] Date in the format of YY-MM-DD.
   \par\noindent
   \footnotemark[$\dagger$] Time in the format of hh:mm.
   \par\noindent
   \footnotemark[$\ddagger$] Net exposure time.
   }\hss}}
\end{tabular}
 \end{center}
\end{table}
%

\section{Analysis}\label{s4}

\subsection{IRSF}\label{s4-1}

\subsubsection{Photometry}\label{s4-1-1}

%
\begin{figure*}
 \begin{center}
  \FigureFile(160mm,80mm){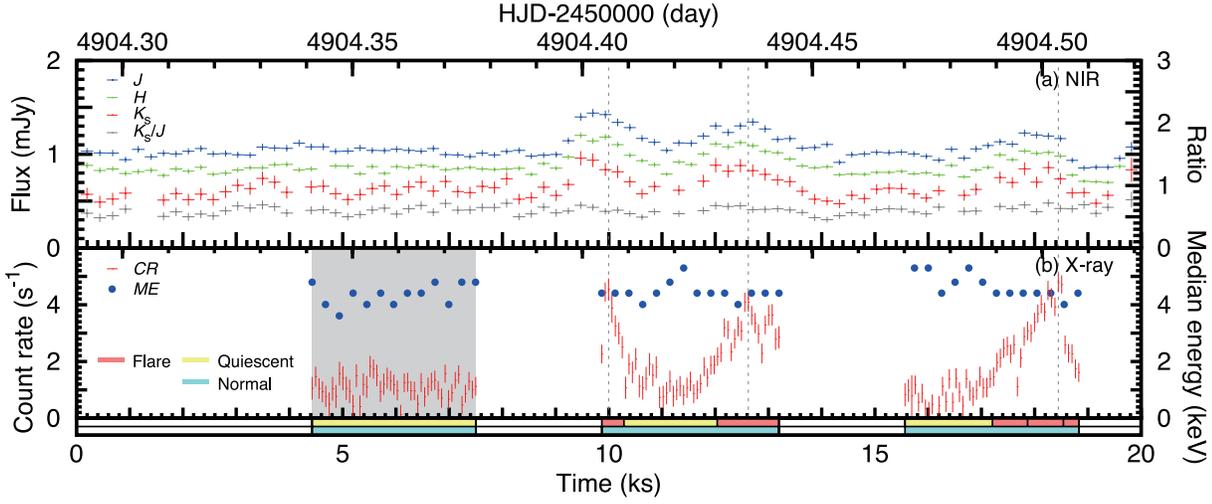}
 \end{center}
 \caption{Simultaneous NIR (SIRIUS) and X-ray light curves on 2009 March 13. 
 The time is shown as Heliocentric Julian Date (HJD) on the top axis and in ks on the bottom axis.
 The origin of the abscissa is 4904.29 in HJD$-2450000$. 
 (a)~NIR light curves. The fluxes in $J$, $H$, and $K_{\mathrm{s}}$ bands (left axis) and the
 color defined as $K_{\mathrm{s}}/J$ (right axis) are respectively represented by blue, green,
 red, and gray cross points.  A typical time bin is $\sim$230~s. 
 (b)~X-ray light curves. The background-subtracted count rate (\textit{CR}; left axis)
 and median energy (\textit{ME}; right axis) are represented by red and blue circles,
 respectively. The \textit{CR} is binned by 64~s, while the \textit{ME} is binned by
 256~s. The 1$\sigma$ Poisson error is shown for the \textit{CR}.
 The mean and standard deviation of \textit{CR} and \textit{ME} are derived from a
 quiescent interval with a gray shaded region. The Bayesian segments and their
 classifications are color-coded in the two horizontal bars at the bottom; 
 flare (red) and quiescent (yellow) for the \textit{CR} segments and normal (blue) for
 the \textit{ME} segments.
 The vertical broken lines indicate the time of X-ray maximum in each flare.}
 \label{f1}
\end{figure*}

We constructed NIR light curves using both SIRIUS and SIRPOL observations. J12270 was
too faint to be detected in each frame exposed for 15~s, thus the minimum time
resolution of the light curves is the time to take a set of 10 dithered frames. For the
SIRPOL data, the Stokes $I$ images were used. Because one $I$ image requires four images
at different position angles, the temporal resolution decreases by 1/4 plus overheads
from the SIRIUS data ($\sim$230~s) to the SIRPOL data ($\sim$1355~s).

The astrometry was calibrated using 28 2MASS \citep{skrutskie06} sources in the FoV,
which are bright but not saturated in SIRIUS images. They are within a magnitude range
of 12.0--15.5, 12.0--15.0, and 12.0--14.0~mag respectively in the $J$, $H$, and
$K_{\mathrm{s}}$ band and with a magnitude error of $\le$0.05~mag.

Aperture photometry was performed with a 2\farcs72 (6~pixel) radius for the source
and a 4\farcs53--9\farcs05 (10--20~pixel) annulus for the sky. The sources used as
astrometric references were also used as relative photometric references, except that we
excluded up to two sources showing a significant deviation from the phenomenological
linear relation between 2MASS and SIRIUS or SIRPOL magnitudes in each band.

\begin{figure*}
 \begin{center}
  \FigureFile(160mm,80mm){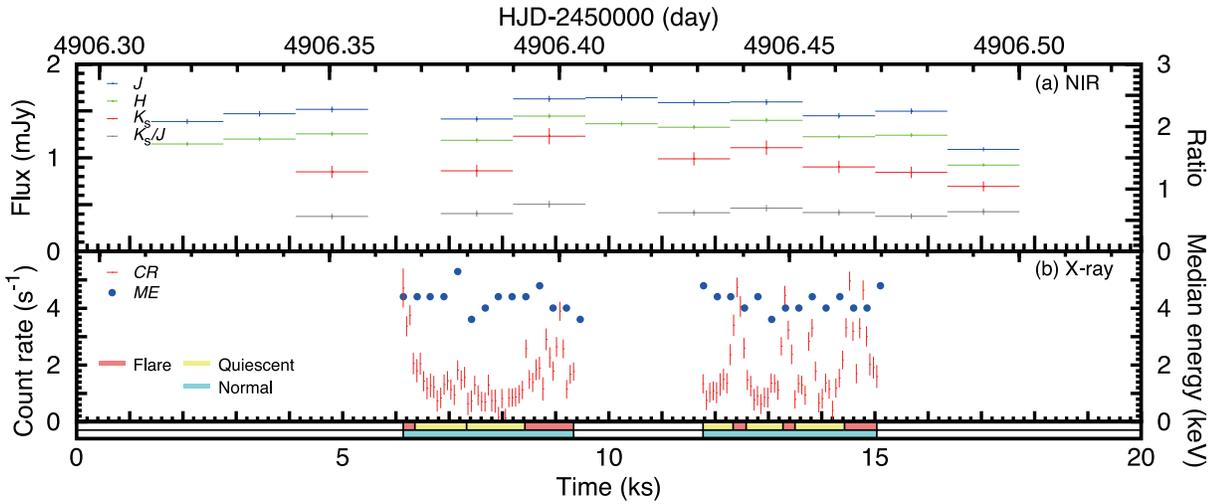}
 \end{center}
 \caption{Simultaneous NIR (SIRPOL) and X-ray light curves obtained on 2009 March
 15. The origin of the abscissa is 4906.295 in HJD$-2450000$.  A typical time bin of
 the NIR data is $\sim$1355~s. Markings follow figure~\ref{f1}.}
 \label{f2}
\end{figure*}

\begin{figure*}
 \begin{center}
  \FigureFile(160mm,80mm){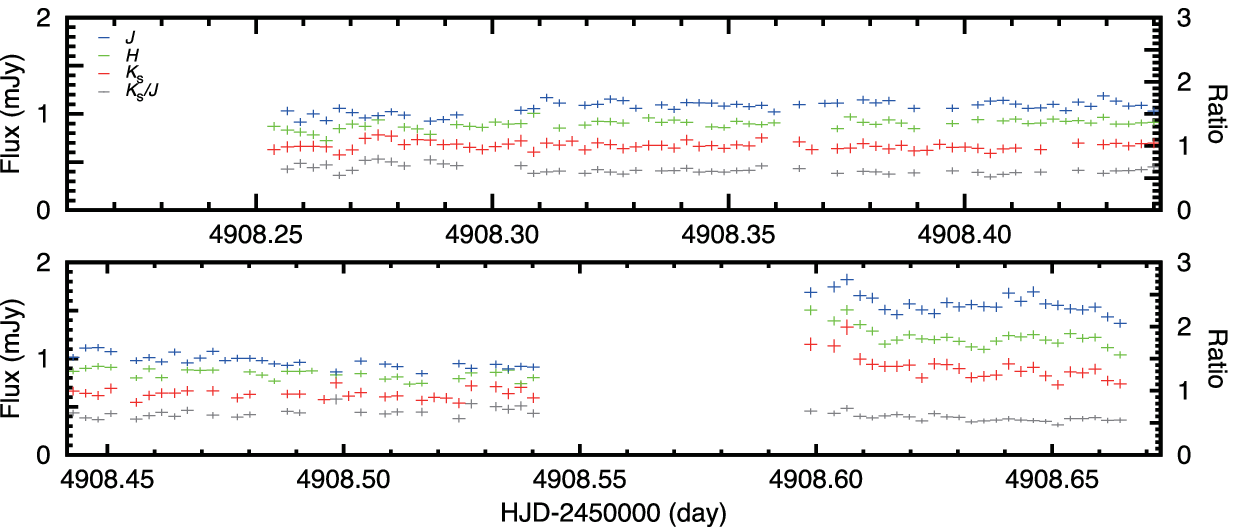}
 \end{center}
 \caption{NIR (SIRIUS) light curves obtained on 2009 March 17. The curves are folded by
 20~ks. The origin of the abscissa is 4908.21 in HJD$-2450000$. Markings follow
 figure~\ref{f1}. A part of the observing time was lost in the latter half due to high
 humidity.}
 \label{f3}
\end{figure*}

Figures~\ref{f1}a, \ref{f2}a, and \ref{f3} show the NIR flux ($J$, $H$, and
$K_{\mathrm{s}}$) and color ($J$--$K_{\mathrm{s}}$) light curves on March 13 (SIRIUS),
15 (SIRPOL), and 17 (SIRIUS), respectively. For the first two figures, the simultaneous
X-ray light curves (\S~\ref{s4-2}) are displayed together. Several flares were clearly
detected in conjunction with X-ray flares. Using all the SIRIUS data sets in
table~\ref{t1}, we constructed a folded light curve at a claimed orbital period of
4.32~hr \citep{demartino10}. However, we could not confirm the period in the NIR bands.

%

\subsubsection{Polarimetry}\label{s4-1-2}
For the SIRPOL data, we performed aperture photometry for images taken at four
different position angles following the same procedure with the SIRIUS photometry. We
then calculated the Stokes parameters $I$, $Q$, and $U$, and derived the polarization
degree $P$. The uncertainties of the Stokes parameters and the polarization degree were
derived by propagating the photometric uncertainty of each source in images at each
position angle. Systematic offset in the polarization degree, which includes both
instrumental artifacts and the global mean of the interstellar dust polarization in the
direction of the object field, were derived (2.2--3.8\%) and corrected using 14 2MASS
sources within $\pm$0.5~mag of J12270 in each band.

\begin{figure}
 \begin{center}
  \FigureFile(80mm,80mm){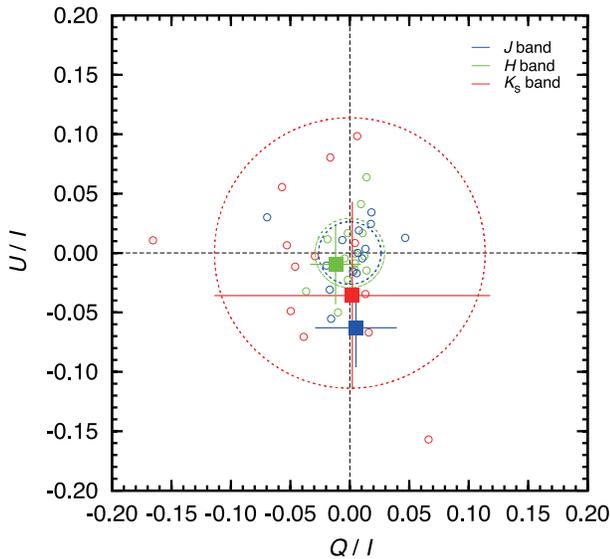}
 \end{center}
 \caption{$Q$/$I$ versus $U$/$I$ plot after offset correction. Large filled squares indicate
 J12270, while small open circles indicate comparison sources within $\pm$0.5~mag of
 J12270. For J12270, the 1$\sigma$ statistical uncertainty is shown with error bars. The
 dotted circles indicate the standard deviation among the comparison sources,
 representing the systematic uncertainty of the origin. Different bands are represented
 by different colors.}
 \label{f4}
\end{figure}

Figure~\ref{f4} shows the $Q$/$I$ versus $U$/$I$ plot of J12270 and the comparison
sources. All SIRPOL data (table~\ref{t1}) were stacked. Assuming that the comparison
sources have no intrinsic linear polarization, the statistical scatter was derived
(dotted circles), which represents the uncertainty in the offset correction in each
band.

We consider that a significant linear polarization was not detected from J12270 in all
bands, because (1)~the polarization degree (distance from the origin in figure~\ref{f4})
is within a 2$\sigma$ convolved uncertainty of the statistical uncertainty in
photometry and the systematic uncertainty in the offset correction and (2)~the
polarization angle (the angle between the $x$-axis and the line connecting the origin
and the data point) is not aligned in the three bands. A 1$\sigma$ upper limit in the
polarization degree was $11.4$, $4.2$, and $12.7$\% respectively in the $J$, $H$, and
$K_{\mathrm{s}}$ bands. A significant polarization was not detected even if the data
were limited only to those during flares.

\subsection{RXTE}\label{s4-2}

\subsubsection{Light Curves}\label{s4-2-1}
We constructed the X-ray light curve in the 2--10~keV band. The background rate was
estimated using the PCA background model provided by the instrument team\footnote{See
http://heasarc.gsfc.nasa.gov/docs/xte/recipes/pcabackest.html for details.}.  Figures
\ref{f1}b and \ref{f2}b show the background-subtracted count rate (\textit{CR}) and the
median energy (\textit{ME}) respectively for the observations in 2009 March 13 and
15. The former observation overlaps partially with a SIRIUS observation, while the
latter with a SIRPOL observation. The \textit{ME} is a proxy for the conventional
hardness ratio \citep{hong04}, which is better suited for low photon statistics and is
defined as the median energy of all photon in a time interval\footnote{Strictly
speaking, we calculated the median of the instrument channel and converted it to median
energy using the energy-channel conversion table. See
http://heasarc.gsfc.nasa.gov/docs/xte/e-c\_table.html for details.}. The \textit{CR} and
the \textit{ME} were binned with 64 and 256~s~bin$^{-1}$, respectively.

Similarly to the Suzaku and XMM-Newton light curves \citep{saitou09,demartino10}, the
RXTE light curves also show various types of variability both in \textit{CR} and
\textit{ME}. We follow \citet{saitou09} for a quantitative assessment of the
variation. First, we divided the \textit{CR} and \textit{ME} curves into ``segments''
with a constant value using the Bayesian blocks method \citep{scargle98}. Second, we
calculated the mean and the standard deviation of \textit{CR} and \textit{ME} in a
featureless interval (the shaded region in figure~\ref{f1}b) to characterize the base
levels. The mean and the standard deviation of the \textit{CR} are 1.07 and
0.37~s$^{-1}$, respectively, while those of the \textit{ME} are 4.34 and 0.39~keV,
respectively. Finally, we classified the segments as (a)~``dips'' for segments with
\textit{CR} below the mean by more than 2$\sigma$, (b)~``flares'' for those with
\textit{CR} above the mean by more than 2$\sigma$, and (c)~``quiescence'' for the
remainder in the \textit{CR} segments, and (i)~``hard'' phase for those with
\textit{ME} above the mean by more than 2$\sigma$ and (ii)~``normal'' for the remainder 
in the \textit{ME} segments. The segments are color-coded following the classification
in figures~\ref{f1}b and \ref{f2}b. We found some flares in both observations, but no
dip and hard phases.

\subsubsection{Spectra}\label{s4-2-2}
\begin{figure}
 \begin{center}
  \FigureFile(80mm,50mm){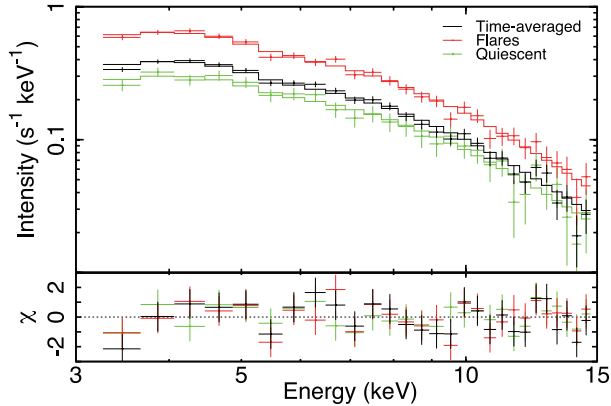}
 \end{center}
 \caption{RXTE background-subtracted spectra of all data (black) together with those
 during flare (red) and quiescent (green) phases. The upper panel shows the data (cross)
 and the best-fit model (solid line), while the lower panel shows the residuals to the
 fit.}
 \label{f5}
\end{figure}

We constructed background-subtracted spectra using all data, those only during the flare
segments, and those only during the quiescent segments (figure~\ref{f5}).  The
background spectra were generated using the same method as for the temporal analysis
(\S~\ref{s4-2-1}). No prominent feature was found. We fitted the 3--15~keV spectra with
a power-law model attenuated by an interstellar absorption model (\texttt{tbabs}:
\cite{wilms00}). We fixed the absorption column density to the value $1.0\times
10^{21}$~cm$^{-2}$ derived from the Suzaku \citep{saitou09} and the XMM-Newton
\citep{demartino10} spectra, as the lack of sensitivity below 2~keV in RXTE did not
allow us to constrain the value. We obtained acceptable fits for the three spectra
(table~\ref{t2}). The power-law index is almost compatible among the three, indicating
that the differences are almost due to the changing flux.

\begin{table}
 \begin{center}
  \caption{Best-fit spectral parameters.\footnotemark[$*$]}
  \label{t2}
  \begin{tabular}{lccc}
   \hline
   State & $\Gamma$\footnotemark[$\dagger$] & $F_{\mathrm{X}}$\footnotemark[$\ddagger$]
   & $\chi^{2}_{\mathrm{red}}$ (dof)\footnotemark[$\S$] \\
   \hline
   Time-averaged \dotfill
      & $1.71^{+0.06}_{-0.06}$
      & $2.45^{+0.06}_{-0.06}$
      & 1.06 (26) \\
   Flares \dotfill
      & $1.77^{+0.07}_{-0.07}$
      & $3.94^{+0.11}_{-0.11}$
      & 0.83 (26) \\ 
    Quiescent \dotfill 
      & $1.63^{+0.13}_{-0.13}$ 
      & $2.00^{+0.10}_{-0.11}$ 
      & 0.52 (26) \\ 
    \hline
   \multicolumn{4}{@{}l@{}}{\hbox to 0pt{\parbox{85mm}{
   \footnotesize
   \par\noindent
   \footnotemark[$*$] The errors are for 90\% statistical uncertainty.
   \par\noindent
   \footnotemark[$\dagger$] Photon index of the power-law.
   \par\noindent
   \footnotemark[$\ddagger$] The observed 3--15~keV flux in units of 10$^{-11}$ erg~s$^{-1}$~cm$^{-2}$.
   \par\noindent
   \footnotemark[$\S$] The reduced $\chi^{2}$ ($\chi^{2}_{\mathrm{red}}$) and the
   degrees of freedom (dof). 
   }
   \hss}}
  \end{tabular}
 \end{center}
\end{table}
%

\subsection{Swift}\label{s4-3}

\subsubsection{Light Curves}\label{s4-3-1}
In the XRT images, the source events were extracted from a circle of a 47\farcs1
(20~pixel) radius covering 90\% of 1.5~keV photons for a point source, while the
background events were taken from an annulus with inner and outer radii of 94\farcs3 (40~pixel) 
and 176\farcs8 (75~pixel), respectively. Some data were affected by photon
pile-up, for which we excluded the central 7\farcs1 circle (3~pixel) and rescaled the
flux to compensate for the reduced aperture \citep{pagani06}.

\begin{figure*}
 \begin{center}
  \FigureFile(160mm,80mm){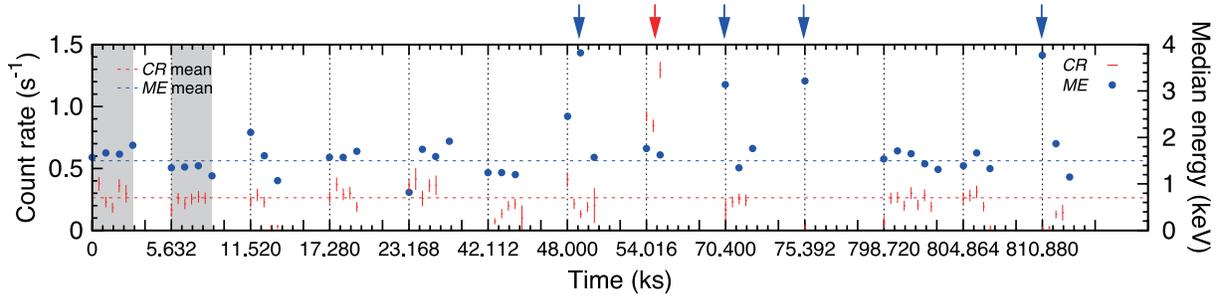}
 \end{center}
 \caption{Swift light curves of the \textit{CR} and the \textit{ME}. Markings follow
 figure~\ref{f1}. The origin of the time represents 3628.51409~d in HJD$-2450000$. Since
 the Swift observations are short and discontinuous, we only present the intervals with
 data and their start time. Red and blue arrows on the top axis indicate possible flares
 and hard phases, respectively. The horizontal broken lines represent the mean of the
 \textit{CR} and the \textit{ME} derived from the data in the shaded intervals.}
 \label{f6}
\end{figure*}

Similarly to the RXTE data, we constructed the \textit{CR} and \textit{ME} light curves
in 0.2--10~keV (figure~\ref{f6}). The \textit{CR} and the \textit{ME} were binned with
128 and 256~s~bin$^{-1}$, respectively. We derived the mean and the standard deviation
of \textit{CR} as 0.26 $\pm$ 0.06~s$^{-1}$, and those of \textit{ME} as 1.49 $\pm$
0.21~keV, respectively, using featureless intervals (the shaded regions in
figure~\ref{f6}). Swift data are composed of 13 snapshots of a very short exposure, so we
did not employ the Bayesian blocks method for segmentation and classification of the
variability. However, we do see some features similar to flares and hard phases as
indicated respectively by red and blue arrows in figure~\ref{f6}.

%

\section{Discussion}\label{s5}

\subsection{NIR Flares}\label{s5-1}
Figure~\ref{f1} convincingly shows that all X-ray flares accompany NIR flares. Figure~2
in \citet{demartino10} also shows that all X-ray flares accompany optical flares. It is
of little doubt that all these phenomena are related to each other.

Looking closely into the flares, it is interesting to point out the following: (1)~NIR
flare peaks appear before the corresponding X-ray peak by $\sim 150-300$~s (shown by
dotted lines in figure~\ref{f1}). 
The optical flare peaks are also seen before the X-ray peak on a similar time scale 
\citep{demartino10}. (2)~No clear change
in NIR color is seen during the flares. The $J$--$K_{\mathrm{s}}$ curve in
figure~\ref{f1} is consistent with being constant in a $\chi^2$ test. (3)~The
amplification from the quiescent level is roughly the same among all NIR flares, which
is also the case for X-ray flares.

Apart from the flux changes associated with X-ray flares, there is also a change in NIR
flux by 0.5~mag between the third and fourth quarters in figure~\ref{f3}. In comparison
to other sources used as photometric reference, we found that this phenomenon is unique
to J12270. It is either that the quiescent base level has changed or that the fourth
quarter was observed during a decay phase of a long flare.

\begin{figure}
 \begin{center}
  \FigureFile(80mm,50mm){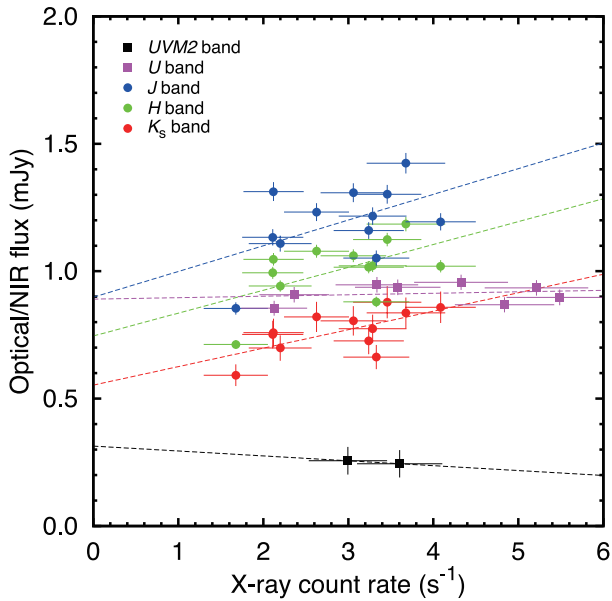}
 \end{center}
 \caption{X-ray count rate (2--10~keV) versus NIR ($J$, $H$, and $K_{\textrm{s}}$) or
 optical ($U$ and $UVM2$) flux during flares. Different bands in the NIR and optical are
 shown in different colors. The dotted lines are the linear regression of the
 relation. The simultaneous RXTE X-ray and NIR data (this work) and XMM-Newton X-ray and
 optical data \citep{demartino10} are used.}
 \label{f7}
\end{figure}

In order to separate the variable and non-variable components in the NIR, we
plotted X-ray count rate versus NIR flux in the flare segments separately for the $J$,
$H$, and $K_{\mathrm{s}}$ bands in figure~\ref{f7}. We used a larger time bin of 300~s
to accommodate a possible time lag between the X-ray and NIR flares. The X-ray and NIR
fluxes are roughly proportional, yielding acceptable fits by a linear relation for the
three bands. The slopes of the three linear relations are consistent with each other,
which is consistent with the NIR flares being colorless. Assuming that the y-intercept
of the linear relations is the residual non-variable component in the NIR flux, it is
estimated as 0.90, 0.75, and 0.55~mJy in $J$, $H$, and $K_{\mathrm{s}}$ band,
respectively. The contamination to the X-ray flux by other sources (\S~\ref{s3-2})
changes this estimate by $\lesssim$10\%.

When we add the optical data taken from \citet{demartino10} in figure~\ref{f7}, a
different and nearly flat slope is found for the optical photometry against the X-ray
count rate during the flare. This might indicate that there are two components in the
NIR and optical bands; one is non-variable during flares and is more intense in the
optical band, while the other is variable during flares with a proportional behavior to
X-ray counts and is more intense in the NIR band.

A plausible origin of the non-variable component is the emission from the secondary star
and a part of the accretion disk, if J12270 is indeed a LMXB. If we assume that the
non-variable NIR flux is entirely from the secondary, we can give a crude constraint to
the spectral type of the companion. By dereddening the NIR color using the X-ray
absorption $N_{\mathrm{H}}=1.0 \times 10^{21}$~cm$^{-2}$ and the conversion ratio of
$N_{\rm{H}}$/$A_{V}=$1.79$\times$10$^{21}$~cm$^{-2}$~mag$^{-1}$ \citep{predehl95}, we
estimate the spectral type of the secondary to be F5\,V--F7\,V if it is a dwarf and
K2\,III if it is a giant \citep{tokunaga00}. The distance is then estimated as
$\sim$1.6~kpc (dwarf) or $\sim$18~kpc (giant). The distance for the giant is too large,
putting this source at the edge or beyond our Galaxy. Therefore, it is more likely that
the secondary is a red dwarf. \citet{demartino10} derived the secondary mass as
$M_{2}=$0.3--0.4~$M_{\odot}$ independently using an empirical dynamical relation between
the orbital period and $M_{2}$, which is slightly smaller than our estimate of $\sim$1.4\MO. 
In any
case, the spectral type determined only from NIR flux should be taken with caution. 
NIR spectroscopic observations are required to decompose the possible secondary
and disk emission and to derive on accurate spectral type of the secondary (e.g.,
\cite{bandyopadhyay97}).

Because the distance is so poorly constrained with the currently available data set, we
normalize physical values with a 1~kpc distance hereafter. The distance does not
contradict the observed $N_{\rm{H}}$ value and its Galactic position. The height from
the Galactic plane would be 240~pc at a 1~kpc distance, which is not exceptionally large
considering that some LMXBs lie beyond this height (e.g., XTE\,J1118+480 at 1.5~kpc;
\cite{mikolajewska05,gelino06}).  An extrapolated X-ray luminosity using the RXTE result
(table~\ref{t2}) is $L_{\mathrm{X\,(0.1-100~keV)}}\sim 1\times 10^{34}\left(
d/1~\mathrm{kpc} \right) ^{2}$~erg~s$^{-1}$.  Combining a $\gamma$-ray luminosity of
$L_{\gamma\,\mathrm{(0.1-300~GeV)}}\sim 5\times 10^{33}\left( d/1~\mathrm{kpc} \right)
^{2}$~erg~s$^{-1}$ (\cite{hill11}), we derived the bolometric luminosity as an order of
$\sim 10^{34}\left( d/1~\mathrm{kpc} \right) ^{2}$~erg~s$^{-1}$, which means that this
source is in a very low luminosity state with $\approx$10$^{-4}$ of the Eddington
luminosity for a stellar mass black hole or neutron star.

\subsection{NIR Polarization}\label{s5-2}
Some LMXBs, particularly black hole binaries in a low luminosity state, are often
accompanied by relativistic jets \citep{fender06}. Synchrotron radiation is emitted over
a wide spectral range. Some evidence suggests that a break occurs between the
optically-thin and optically-thick synchrotron emission around the NIR band
(e.g. \cite{fender06}). It is also expected that the contrast of the synchrotron
emission is relatively high against other spectral component in this band. Therefore,
many efforts have been made to detect linear NIR polarization from many LMXBs
\citep{dubus06,shahbaz08,russell08}, which would argue for synchrotron origin of the NIR
emission.

However, only two sources (Sco\,X-1 and GRO\,J1655--40; \cite{russell08}) were detected
with NIR polarization intrinsic to the source. The polarization degree was a few percent
at most, which is smaller than our upper limit for J12270 (\S~\ref{s4-1-2}). Our
observation was not sensitive enough to probe possible linear polarization, and the lack
of detection should not discourage further investigation.

\subsection{Long-term X-ray Behavior}\label{s5-3}
We have looked into all the X-ray data available to date through \citet{butters08},
\citet{saitou09}, \citet{demartino10}, and this work using the Suzaku, XMM-Newton, RXTE,
and Swift satellites over a span of 3.5~years starting in 2007 November with a total
integration of 133~ks. The anomalous temporal behavior is seen in all the data sets. The
best-fit parameters in the X-ray spectral fitting during flares and those outside of
flares are almost the same in all data. Also, light curves in several X-ray bands are
available from 2009 August up to the time of writing, from the Monitor of All-sky X-ray Imager
onboard the International Space Station\footnote{See
http://maxi.riken.jp/top/maxi\_data/star\_data/J1227-488/J1227-488\_00055058g\_lc.png
for the latest data.}, which are stable with no apparent phase changes seen in other
monitored LMXBs. All these suggest that the X-ray characteristics presented in previous
and this work are the norm of this object.

\citet{hill11} claim that the Fermi source, for which J12270 is the most likely
counterpart, is a millisecond pulsar, although the optical observation conducted in 2008
July \citep{pretorius09} clearly shows a signature of accretion. To reconcile this
inconsistency, they speculated that the system changed from accretion-powered phase to
rotation-powered phase before the Fermi observation in 2008 August to 2010
September. However, we consider this to be unlikely because the X-ray results throughout
this period show no significant changes in its spectral and temporal characteristics of
the source.

\subsection{X-ray Flares}\label{s5-4}
%
\begin{figure*}
 \begin{center}
  \FigureFile(170mm,170mm){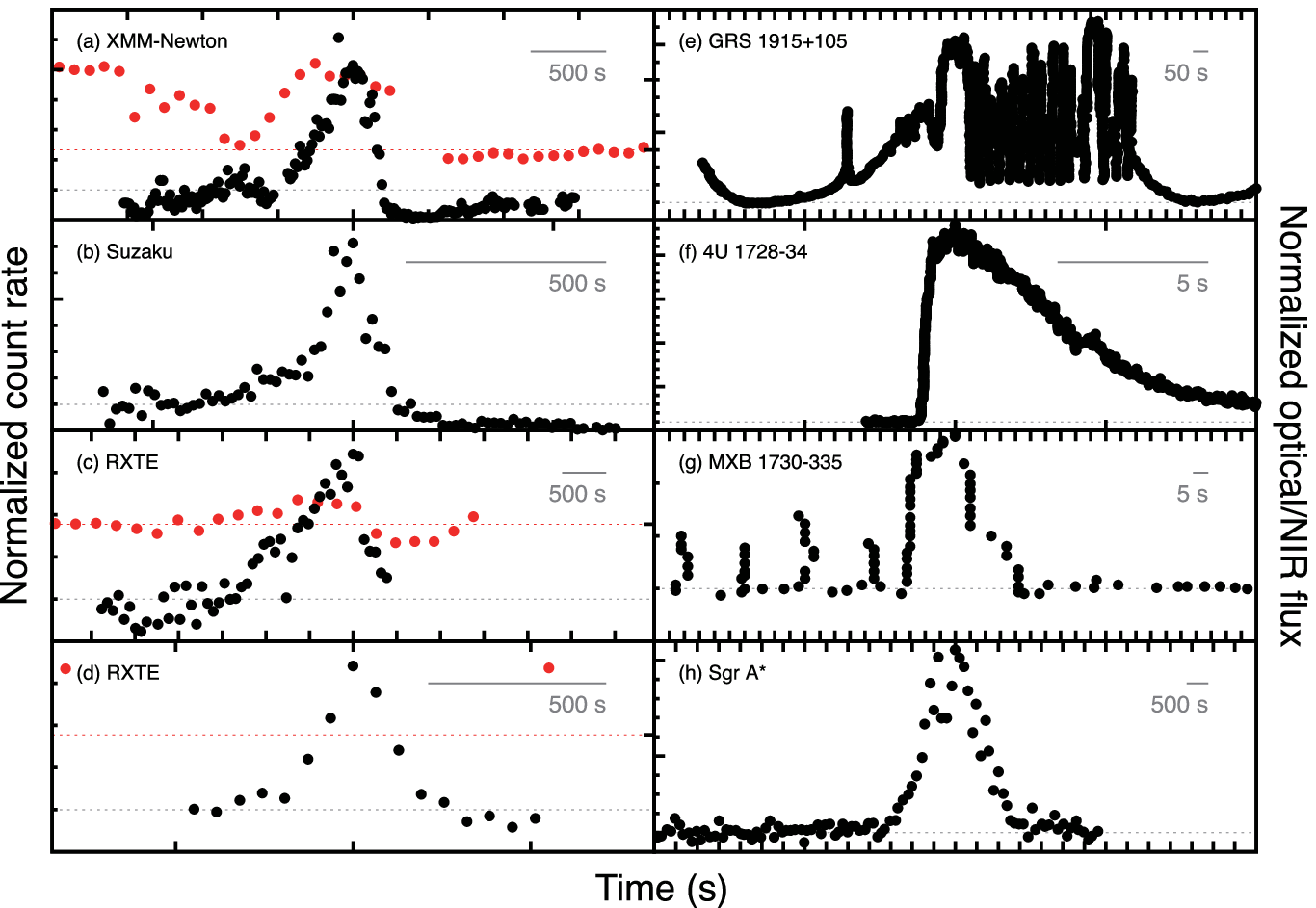}
 \end{center}
 \caption{Flare profiles of (a--d) J12270 in comparison with those in other X-ray
 sources showing repetitive flares: (e)~a microquasar (GRS\,1915$+$105 in the class
 $\beta$ variation; \cite{belloni00}), (f)~a type~I burster (4U\,1728--34;
 \cite{vanstraaten01}), (g)~a type~II burster (MXB\,1730--335 a.k.a. the Rapid Burster;
 \cite{lewin76}), and (h)~Sgr\,A* \citep{porquet03}. The flare samples of J12270 are
 taken at (a)~4837.19, (b)~4687.82, (c)~4904.50, and (d)~4906.44 in HJD$-$2450000.
 Black points represent the X-ray count rate for all sources, while red circles
 represent optical/NIR flux for J12270.
 The scales are arbitrarily expanded to facilitate comparison. The abscissae are ticked
 by 500~s for (a--d, h), by 50~s for (e), and by 5~s for (f--g), while the ordinates are 
 ticked with the quiescent level of being unity.
 The quiescent level, shown with the horizontal dotted line, is measured from a
 non-flaring period.}
 \label{f8}
\end{figure*}

With a close inspection of X-ray flares, some features are seen commonly in all events:
(1)~a time profile of slow rise and rapid decay, (2)~the constant contrast between the
flare peak and the base level, (3)~no spectral changes during flares, (4)~undershoot in
flux and spectral hardening after flares, which are most notably seen in the last flare
in a series of repetitive flares, and (5)~simultaneous optical and NIR flares and their
flare peaks before an X-ray flare peaks.

A difference among these flares is the time scale of the profile. To test this idea, we
used one of the XMM-Newton flares \citep{demartino10} as a template and fitted all the
other X-ray flares by thawing the time scale. Figure~\ref{f8}~(a--d) shows the result,
in which we present the template flare profile (a) and three other flares (b--d) scaled
with the best-fit time scaling factor. We found that the profiles are similar to each
other, including the development in the spectral hardness and optical/NIR flux.

The repetitive nature of such X-ray flares is a distinctive character of J12270. We
compare the flare profile with other classes of Galactic sources showing repetitive
flares and having either a black hole or a neutron star in the system
(figure~\ref{f8}). We pick up representative sources from each class;
(e)~GRS\,1915$+$105 for microquasars (\cite{belloni00}), (f)~4U\,1728--34 for type~I
bursters (\cite{vanstraaten01}), (g)~MXB\,1730--335 for type~II bursters
(\cite{lewin76}), and (h)~Sgr\,A* (\cite{porquet03}). GRS\,1915$+$105 is known to show a
wide variety of flare phenomenology. We compared with the class $\beta$ variation in a
phenomenological classification scheme by \citet{belloni00}.

\citet{demartino10} argued that J12270 is unlikely to be a type~I and type~II
burster. Type~I bursters are characterized by rapid flux increase and slow exponential
decay in their flare profile (figure~\ref{f8}f) and show clear X-ray spectral changes
during flares. Type~II bursters show a correlation between the flare energy and the
waiting time for the flare. Both of these defining characteristics are not seen in
J12270, ruling out the possibility that J12270 is either type~I or type~II bursters.

We consider that the similarities between J12270 and GRS\,1915$+$105 (class $\beta$ 
variation in \cite{belloni00}) may be more relevant for unveiling the nature of this source. 
Both show repetitive flares in similar shapes but with slightly varying time scales,
simultaneous NIR flares \citep{eikenberry98}, no spectral changes during flares, flux
undershoot and spectral hardening after flares, most noticeably in the last flare in a
series, and dichotomy between time intervals filled with flares and those without. There
are some differences; e.g., a precursor flare is seen before a series of repetitive
flares in GRS\,1915$+$105, while no such feature is apparent in J12270. Also,
GRS\,1915$+$105 is known to have lots of other states \citep{belloni00}, while J12270 is
known to have only one so far. GRS\,1915$+$105 is close to the Eddington luminosity
while J12270 is probably at very low luminosity. Despite these differences, it is
interesting that many distinctive features of J12270 are also seen in
GRS\,1915$+$105.

One interesting source that may bridge between the two is IGR\,J17091--3624 
\citep{altamirano11a,altamirano11b}. This source also shows repetitive flares in a
similar time scale and a similar amplitude. The variation is reminiscent of the $\beta$
class variation in GRS\,1915$+$105 \citep{belloni00}, but the luminosity is nowhere near
the Eddington limit. This may indicate that the anomalous flux variation seen in the
three samples can occur at any luminosity.

Based on this, we speculate that the underlying physics to cause such X-ray variability
is the same. The cause for the variability is not well understood for GRS\,1915$+$105,
let alone J12270. At least, we speculate two things: (1)~The NIR flare peaks 150--300~s
before X-ray peaks excludes the possibility of X-ray reprocessing to be the origin of
the NIR flares. (2)~The time scale of the flares in J12270 is much longer than the
scales that are considered to be caused by jet ejection in other microquasars (e.g.,
GX\,339--4 and XTE\,J1118+480; \cite{casella10,gandhi10,kanbach01}). This suggests that
the X-ray and NIR variability is not caused by jet ejection itself, but some sort of
outside-in instability in the accretion disk or an extended corona. In fact, the flare
peak delay between NIR and X-ray (\S\ref{s5-1}) is comparable with the dynamical time
scale of $\sim$150~s for a 5~$M_{\odot}$ black hole at a 10$^{4}$ $r_{\mathrm{g}}$ or
$\sim$170~s for a 1.4~$M_{\odot}$ compact object at a 2.5$\times 10^{4}$
$r_{\mathrm{g}}$, in which $r_{\mathrm{g}}$ is the gravitational radii.

\subsection{Spectral Energy Distribution}\label{s5-5}
\begin{figure*}
 \begin{center}
  \FigureFile(140mm,90mm){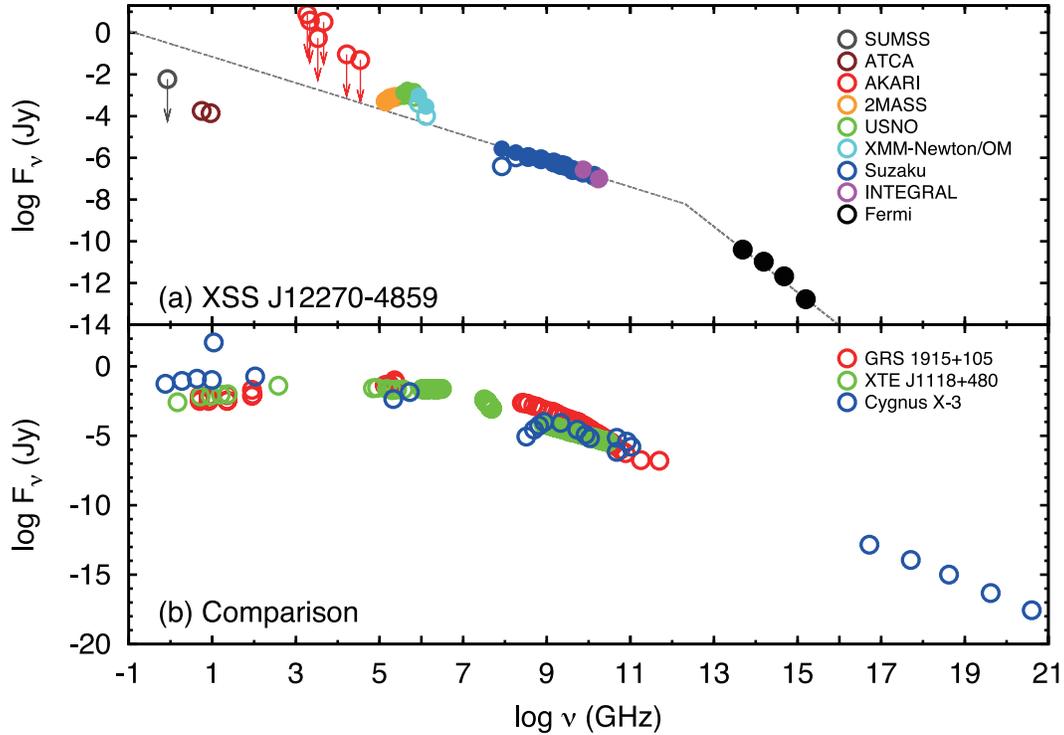}
 \end{center}
 \caption{SED of (a)~J12270 in comparison with (b)~some microquasars. (a)~Open symbols
 indicate observed flux, while filled symbols indicate extinction-corrected flux. For
 the SUMSS and AKARI data, only upper limits are given, which are indicated by downward
 arrows. The broken lines represent extrapolated power-law model derived from the Suzaku
 or Fermi data. (b)~The data are taken from \citet{ueda02} for GRS\,1915$+$105,
 \citet{markoff01} for XTE\,J1118$+$480, and \citet{sinitsyna09} for Cygnus\,X-3
 regardless of their states.}
 \label{f9}
\end{figure*}

It is vital to construct an SED to understand the nature of high-energy objects. To
extend the SED shown in \citet{demartino10} to longer wavelengths, we retrieved survey
data at many wavelengths, which include the Sydney University Molonglo Sky Survey
(SUMSS; \cite{mauch03}) at 843~MHz, the AKARI all-sky survey in far-infrared at 160,
140, 90, and 65~$\mu$m \citep{yamamura10} and in mid-infrared at 18 and 9~$\mu$m
\citep{ishihara10}, the 2MASS all-sky survey catalogue \citep{skrutskie06} in NIR, the
US Naval Observatory catalogue (USNO-B; \cite{monet03}) in the optical, the fourth
IBIS/ISGRI soft gamma-ray survey catalog \citep{bird10}, and the Fermi Large Area
Telescope first source catalogue \citep{abdo10}. We also added the recent ATCA radio
detection at 5.5 and 9.0~GHz by \citet{hill11}. We made an SED for observed as well as
extinction-corrected fluxes using bands with a significant detection (figure~\ref{f9}
top). We found only upper limits for the SUMSS and AKARI data, which are indicated by
downward arrows.

As a comparison, we present the SED of some microquasars: GRS\,1915$+$105,
XTE\,J1118$+$480, and Cygnus\,X-3 (figure~\ref{f9} bottom). In the lower energy band,
J12270 shares a common characteristic with the other microquasors, in which a nearly
flat SED is seen down to $\sim$1~GHz. This is interpreted as optically-thick synchrotron
emission from relativistic jets \citep{blandford79}.

In the upper energy band, the $\gamma$-ray spectrum of J12270 is apparently steeper and
lower than its extrapolated X-ray spectrum. However, we can fit the X-ray to GeV SED
with a simple broken power-law model with a break at $\sim 2\times 10^{12}$~GHz
(figure~\ref{f9}). The break at $\sim$10$^{11}$~GHz for microquasars is expected and
interpreted as the consequence of the synchrotron cooling in the optically-thin regime
of the emission. The break frequency $\nu_{\mathrm{max}}$ is determined from the balance
between the radiative cooling and acceleration of electrons. If Fermi acceleration in
shock waves is responsible for the acceleration,
\begin{eqnarray}
\nu_{\mathrm{max}} = 1.2 \times 10^{11} \left(\frac{\xi}{100}\right)^{-1}
 \left(\frac{v_{\mathrm{shock}}}{c}\right)^{2}~\mathrm{[GHz]},\nonumber
\end{eqnarray}
where $v_{\mathrm{shock}}$ is the shock velocity and $\xi$ is the dimensionless
parameter to represent the turbulence in the magnetic field in the shock; i.e., mean
free path of an electron in units of the Larmor radius \citep{markoff01}.

Due to the break, microquasars are just below the sensitivity of the Fermi survey
\citep{abdo10}. However, with a slight change in the parameters, a Fermi detection is
not a surprise for microquasars. Indeed, the first such detection was reported from
Cygnus X-3 recently by \citet{corbel10}. J12270 shows a break at a frequency higher by a
decade, which may indicate that it has a slightly more efficient acceleration in
comparison to others, if it is a microquasar. 

On the other hand, no AKARI detection and the ATCA detection below back-extrapolating
the X-ray power-law (figure~\ref{f9}) suggest that there is a second break to the
broadband power-law emission. In fact, a break in IR band attributed to the jet is
observed in a microquasar GX\,339--4 (\cite{corbel02}).  Those two breaks at high energy
and IR bands are reported in a black hole candidate XTE\,J1118$+$480 in the low/hard
state (\cite{markoff01}), whose emission is considered to be mainly came from
synchrotron jet.

\section{Summary}
We conducted NIR and X-ray observations of the enigmatic source
XSS\,J12270--4859. During the coordinated NIR and X-ray observations using IRSF 
SIRIUS and RXTE PCA, we detected simultaneous and repetitive NIR and X-ray flares for
the first time. NIR polarization was not detected. Based on the observed data together
with those presented in previous work \citep{saitou09,demartino10}, we compared the
flare profiles and the SED with other classes of X-ray sources and argued that J12270
shows similarities with a type of variation seen in the microquasar GRS\,1915$+$105 in
many of its defining characteristics. Based on this and the broad SED commonly seen in
microquasars, we argued that the nature of J12270 is a microquasar. At an assumed
distance of 1~kpc, it has a very low luminosity of $\approx$10$^{-4}$ of the Eddington
luminosity for a stellar mass black hole or neutron star. If J12270 is confirmed to be a
LMXB with a Fermi counterpart, it will be the first example of a $\gamma$-ray binary
having a low-mass companion.

\bigskip

We thank Daisuke Kato for his help in obtaining NIR data for a SIRIUS run, Hirofumi
Hatano for advice in polarimetry data reduction, Shinki Oyabu for advice on AKARI
survey data, and Hajime Inoue and Mamoru Doi for useful discussion. We appreciate the
telescope managers of RXTE for allocating a telescope time for our
observation. K.~S. is financially supported by the Japan Society for the Promotion of
Science and the Hayakawa foundation of the Astronomical Society of Japan.

This research made use of data obtained from Data ARchives and Transmission System
(DARTS), provided by Center for Science-satellite Operation and Data Archives (C-SODA)
at ISAS/JAXA, and from the High Energy Astrophysics Science Archive Research Center
Online Service, provided by the NASA/Goddard Space Flight Center. We also made use of
the SIMBAD database, operated at CDS, Strasbourg, France. IRAF is distributed by the
National Optical Astronomy Observatories, which are operated by the Association of
Universities for Research in Astronomy, Inc., under cooperative agreement with the
National Science Foundation.


\end{document}